\documentclass[journal]{IEEEtran}
\usepackage{graphicx}
\usepackage{cite}
\usepackage{picinpar}
\usepackage{amsmath}
\usepackage{url}
\usepackage{flushend}
\usepackage[latin1]{inputenc}
\usepackage{colortbl}
\usepackage{soul}
\usepackage{multirow}
\usepackage{pifont}
\usepackage{color}
\usepackage{alltt}
\usepackage[hidelinks]{hyperref}
\usepackage{enumerate}
\usepackage{siunitx}
\usepackage{breakurl}
\usepackage{epstopdf}
\usepackage{pbox}
\usepackage{makecell}
\usepackage{amssymb}
\usepackage{algorithm}
\usepackage{algorithmic}
\ifCLASSINFOpdf
\else
\fi
\begin{document}
%
\title{Cooperative Decision Making of Connected Automated Vehicles at Multi-lane Merging Zone: A Coalitional Game Approach}
%
%
%

\author{Peng Hang, Chen Lv,~\IEEEmembership{Senior Member,~IEEE,}
        Chao Huang, Yang Xing, and Zhongxu Hu
\thanks{This work was supported in part by the A*STAR Grant (No. 1922500046), Singapore and the SUG-NAP Grant (No. M4082268.050) of Nanyang Technological University.}
\thanks{P. Hang, C. Lv, C. Huang, Y. Xing and Z. Hu are with the School of Mechanical and Aerospace Engineering, Nanyang Technological University, Singapore 639798. (e-mail: \{peng.hang, lyuchen, chao.huang, xing.yang, zhongxu.hu\}@ntu.edu.sg)}
\thanks{Corresponding author: C. Lv}}

%
%

\markboth{ }%
{Shell \MakeLowercase{\textit{et al.}}: }
%



\maketitle

\begin{abstract}
To address the safety and efficiency issues of vehicles at multi-lane merging zones, a cooperative decision-making framework is designed for connected automated vehicles (CAVs) using a coalitional game approach. Firstly, a motion prediction module is established based on the simplified single-track vehicle model for enhancing the accuracy and reliability of the decision-making algorithm. Then, the cost function and constraints of the decision making are designed considering multiple performance indexes, i.e. the safety, comfort and efficiency. Besides, in order to realize human-like and personalized smart mobility, different driving characteristics are considered and embedded in the modeling process. Furthermore, four typical coalition models are defined for CAVS at the scenario of a multi-lane merging zone. Then, the coalitional game approach is formulated with model predictive control (MPC) to deal with decision making of CAVs at the defined scenario. Finally, testings are carried out in two cases considering different driving characteristics to evaluate the performance of the developed approach. The testing results show that the proposed coalitional game based method is able to make reasonable decisions and adapt to different driving characteristics for CAVs at the multi-lane merging zone. It guarantees the safety and efficiency of CAVs at the complex dynamic traffic condition, and simultaneously accommodates the objectives of individual vehicles, demonstrating the feasibility and effectiveness of the proposed approach.
\end{abstract}


\IEEEpeerreviewmaketitle

\section{Introduction}
\subsection{Motivation}
\IEEEPARstart{T}{raffic} congestion has become a distressing issue for a long time. It reduces the traffic efficiency, leading to discomfort for passengers, and increasing the energy consumption and the risk of collisions [1]-[3]. Among various causes of traffic congestion, vehicle merging is a typical one, especially in highway scenarios [4]. Generally, a vehicle on the on-ramp lane has to slow down to wait for a desired chance to merge in [5]. Since the velocity of the vehicle on the on-ramp lane is smaller than that of the vehicles on the main lane, it would cause three consequences: (1) The vehicle on the on-ramp lane can conduct merging successfully, but the vehicle on the main lane is forced to decelerate suddenly that affects the traffic efficiency; (2) A vehicle collision occurs during the merging process due to small safe gap or aggressive driving behaviors, which further results in a serious traffic congestion; (3) If the driving pattern of the vehicle on the on-ramp lane is conservative or calm, it would consume more time for the vehicle to complete its merging maneuver, and the traffic efficiency of the on-ramp lane will be also affected. In general, decision making in lane merging scenarios is always not easy, even for human drivers, due to many complex factors, including the dynamic traffic states, the driving preferences and travel objectives of vehicles.

Automated driving and vehicle to vehicle (V2V) technique are promising pathways to address the traffic congestion issue and further improve driving safety [6]. Under a connected driving environment, not only the vehicles' dynamic states, but also their driving intentions, behaviors and surrounding environment information can be exchanged and shared. This makes the multi-vehicle interactions under complex environment possible [7]. CAVs are able to make reasonable and effective decisions to address the merging issue cooperatively. Then, the driving performance of the mobility system, including the safety, efficiency, comfort and energy, can be enhanced accordingly [8].

\subsection{Related Work}
To address the lane merging issues for CAVs, the centralized control approach has been widely studied by many researchers [9]-[11]. A centralized controller is proposed to control and manage the vehicles within the certain controlled area [12]. Upon entering the controlled area, CAVs need to hand over their control authorities to the centralized controller. Then, the centralized control unit optimizes the entire traffic sequence, and guides the vehicles on the on-ramp lane to enter the main lane orderly at a defined merging point.

To maximize the travel efficiency of CAVs, a longitudinal freeway merging control algorithm, in which CAVs follow the order assigned by the roadside controller, is proposed [13]. In [14], a broadcast controller is designed considering the pseudo perturbation to coordinate the CAVs on multiple lanes to realize smooth merging behaviors. To reduce the computing time and enhance the coordination performance, a grouping-based cooperative driving approach is studied to address the merging issue of CAVs [15]. In [16], a cooperative ramp merging framework is built for CAVs and human-driven vehicles with a bi-level optimization, which enables the cooperative and noncooperative behaviors in the mixed traffic environment. In addition, the vehicle platooning is an effective way to improve the merging efficiency [17]. In [18], a communication network is designed to control the lane-change actions of CAVs and the merging behaviors of the vehicle platoon. In [19], model predictive control (MPC) is applied to the vehicle platooning control to solve the multi-vehicle merging issue. Moreover, the concept of spring-mass-damper system is also utilized to improve the platoon's travel efficiency and stability after merging [20].

The aforementioned studies mainly focus on optimizing the merging sequence of CAVs (i.e., the longitudinal motion optimization) to advance the efficiency of traffic flow [21, 22]. However, the traffic scenario is mainly limited to a single-lane main road and an on-ramp lane, and the lane-change behavior of the vehicle on the main lane is usually neglected. Besides, the lateral motion optimization of CAVs during merging is rarely studied. In [23], an online control algorithm for vehicles at multi-lane merging zones is proposed via optimizing the lane-change and car-following trajectories of CAVs. To improve the merging efficiency on a multi-lane road, a MPC-based framework is established, which is able to generate the optimal acceleration and make safe lane-change decisions simultaneously [24].

Besides the above approaches, game theory is another effective method to address the decision-making issues of CAVs [25]. In [26], a cooperative game approach is applied to the on-ramp merging control problem of CAVs, which can reduce the fuel consumption and travel time and further improve the ride comfort. In [27], a human-like game theory-based controller is designed for automatic lane change of CAVs. In [28], a game theoretic approach is applied to the predictive control for lane-change and car-following of CAVs. Nevertheless, the game theoretical approaches can be applied not only to handle the decision making of CAVs, but also to mimic the interactive behaviors of intelligent multi-agent.

\subsection{Contribution}
To further advance the safety and smartness of mobility systems, in this paper, a coalitional game approach is developed to address the cooperative decision-making problem for CAVs at multi-lane merging zone with consideration of different interactive driving behaviors. The contributions of this paper are summarized as follows: (1) A cooperative decision making framework is proposed based on the coalitional game theory to deal with the multi-lane merging of CAVs at the multi-lane merging zone. This is to advance the driving performances for CAVs, including the safety, comfort and efficiency; (2) The motion prediction of CAVs is considered within the cooperative merging framework via MPC to enhance the effectiveness of the decision making; (3) The interactions and decision-making of CAVs with different driving characteristics are studied. This yields multi-modal coalitional combinations and decision-making strategies, which can satisfy the personalized demands of CAVs.
\subsection{Paper Organization}
The remainder of the paper is organized as follows. Section II presents the problem formulation and high-level system framework of the decision making for multi-lane merging of CAVs. In Section III, the motion prediction model of CAVs is established. In Section IV, the cooperative decision-making algorithm lane merging of CAVs is designed with the coalitional game approach. Section V presents the testing results and analysis. In Section VI, this study is concluded.

\section{Problem Formulation and System Framework}
\subsection{Problem Formulation}
Although the centralized control approach is able to manage the status of the entire traffic system, the traffic efficiency is not the only consideration for control of CAVs. Different passengers have distinguished demands for their mobilities. For instance, family travellers especially with babies onboard, as well as aged occupants, prefer a more comfortable and safer riding experience. While, commuters, especially during peak hours, care more about their travel efficiencies. Therefore, future automated driving should be designed in personalized and human-like ways. To do this, the driving characteristics or modes can be embedded into the CAVs so as to adapt to different demands from passengers. Thus, in this context, not all vehicles are willing to hand out their control authorities to the centralized controller, instead, many individual users prefer to make their personalized decisions and travel modalities.

\begin{figure}[t]\centering
	\includegraphics[width=8.5cm]{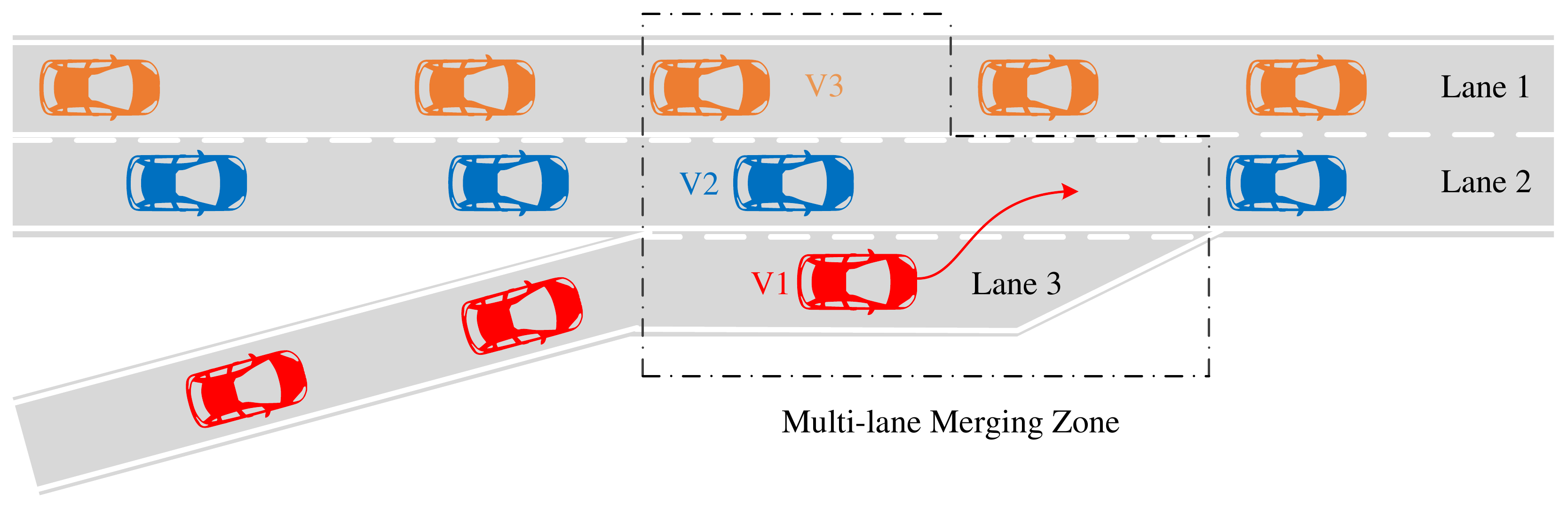}
	\caption{Decision making of CAVs at the multi-lane merging zone.}\label{FIG_1}
\end{figure}

For the centralized traffic control system, many existing studies consider a two-lane merging scenario, which only consists of a single-lane main road and an on-ramp lane. The lane-change maneuver on the main lane is usually neglected. In this paper, to further advance the algorithm and expand the system complexity, a multi-lane merging scenario is investigated. As shown in Fig. 1, the merging scenario on the highway consists of three lanes, i.e., two main lanes and an on-ramp lane. Once V1 enters the multi-lane merging zone from the on-ramp lane, it will need to interact with its surrounding vehicles and make a decision on its merging behavior. In that case, on one hand, V2 can either slow down or change lane to give way to V1. On the other hand, instead of giving way, it may remain its speed or even accelerate to compete for the right of way. Besides, the operation behavior of V2 would also affect the decision making of its adjacent car V3. In general, the decision making and driving behaviors of CAVs within the merging zone are affected by each other with their different driving characteristics and objectives. With the consideration of driving safety, ride comfort and travel efficiency of the system as well as the individual vehicles, this paper investigates the interactions and decision making strategies of the CAVs at the multi-lane merging zone by using a coalitional game approach.

\subsection{Cooperative Decision-Making Framework for CAVs}
As Fig. 2 shows, a cooperative decision-making framework is proposed to address the multi-lane merging issue of CAVs. In this study, three human-like driving characteristics, i.e., aggressive, moderate and conservative, are defined for automated vehicles. The aggressive driving style gives the highest priority to the travel efficiency, therefore, it would lead to aggressive driving maneuvers, such as sudden accelerations or decelerations. The conservative one cares more about safety and comfort, rather than travel efficiency. While the moderate one is positioned in between the aforementioned two categories, expecting a balance among multi-objective of driving performance [29].

The motion prediction module provides the predicted motion states of the vehicles for decision making. According to the predicted motion states of the host CAV and surrounding CAVs, the cost function for decision making is then formulated considering multiple constraints. Based on the cost functions of decision-making for the CAVs, a cooperative game theory approach, i.e., the coalitional game, is applied to deal with the coordination and decision-making for CAVs at the multi-lane merging zone. After solving the formulated optimization problem of the cooperative game in the cloud, the decision-making results are generated and sent to the motion planning module of each CAV. Finally, the motion controller executes the expected decision-making command. Some signals of the motion-control module, for example the vehicles' dynamic sates and positions, are output to the decision-making module, which forms a closed-loop between decision making and motion control. The design process of the motion-planning module with a potential field approach has been reported in a previous work [30]. And in this study, we mainly focus on the algorithm design of the decision making module.

\begin{figure}[!t]\centering
	\includegraphics[width=8.5cm]{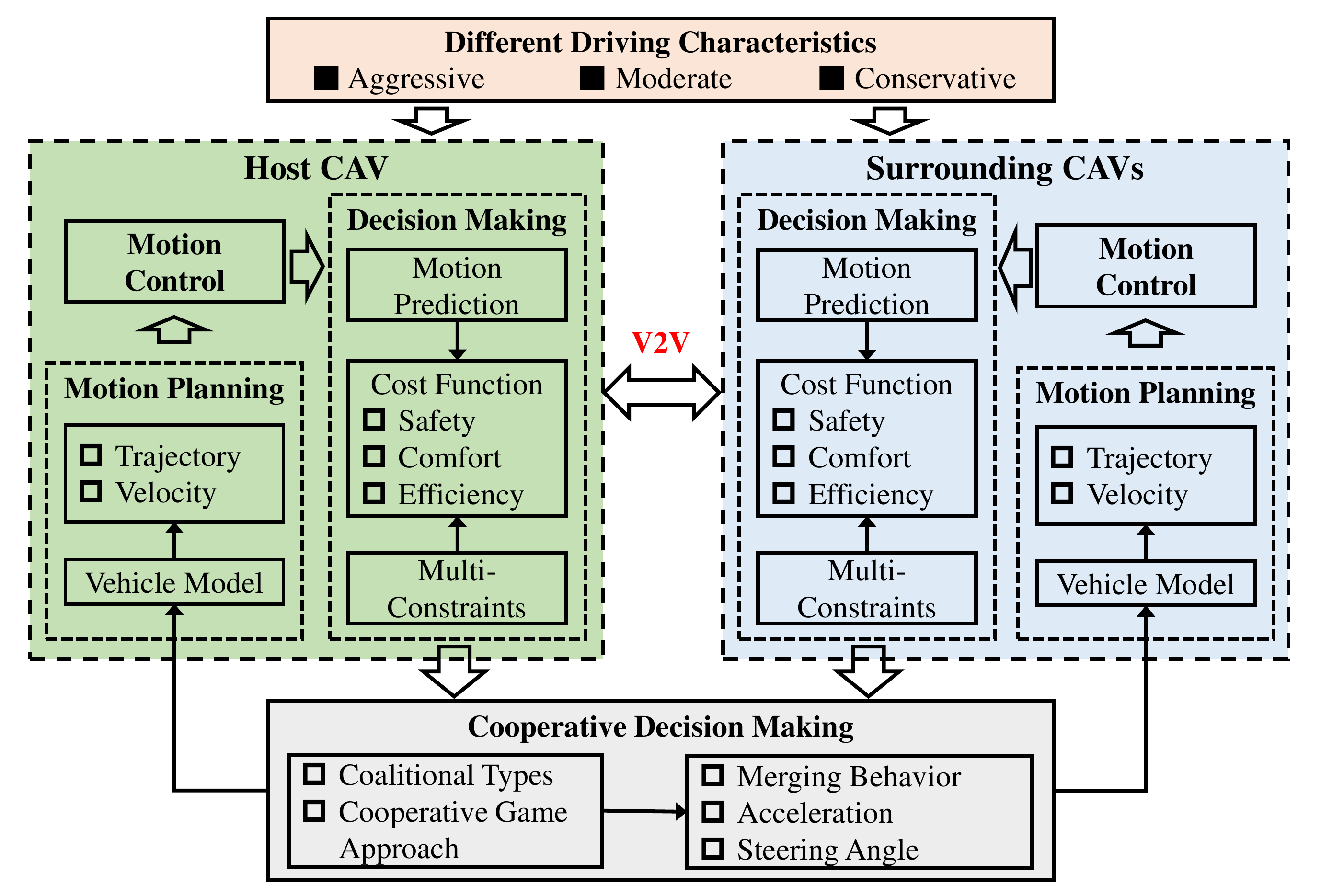}
	\caption{The proposed decision making framework for CAVs.}\label{FIG_2}
\end{figure}

\section{Motion Prediction of CAVs}
In this section, a single-track vehicle model is proposed for the motion prediction of CAVs. Then, based on this single-track vehicle model, a discrete motion prediction model is further established for design of the decision-making algorithm.
\subsection{The Single-track Vehicle Model}
To reduce the model complexity in motion prediction, the four-wheel vehicle model is simplified into a bicycle model. With the assumption of a small steering angle $\delta_f$ at the front wheel, it yields that $\sin\delta_f\approx0$. Then the single-track bicycle model built for motion prediction can be expressed as follows [31-33].
\begin{align}
\dot{x}(t)=\Gamma[x(t),u(t)]
\end{align}
\begin{align}
\Gamma[x(t),u(t)]=
&
\left[
\begin{array}{ccc}
v_yr+F_{xf}\cos\delta_f/m+F_{xr}/m\\
-v_xr+F_{yf}\cos\delta_f/m+F_{yr}/m\\
l_fF_{yf}\cos\delta_f/I_z-l_rF_{yr}/I_z\\
r\\
v_x\cos\varphi-v_y\sin\varphi\\
v_x\sin\varphi+v_y\cos\varphi\\
\end{array}
\right]
\end{align}
where the state vector $x=[v_x,v_y,r,\varphi,X,Y]^{T}$,and the control vector $u=[a_x,\delta_f]^{T}$. $v_x$ and $v_y$ are the longitudinal and lateral velocities, respectively. $r$ and $\varphi$ are the yaw rate and yaw angle, respectively. $(X,Y)$  is the coordinate position of the vehicle. $F_{xi}(i=f,r)$ and $F_{yi}(i=f,r)$ are the longitudinal and lateral tire forces of the front and rear wheels. $l_f$ and $l_r$ are the distances from the center of mass to the front axle and the rear axle, respectively. $m$ is the vehicle mass, and $I_z$ is the yaw moment of inertia.

Neglecting the air resistance and the rolling resistance, the longitudinal dynamics can be further simplified as
\begin{align}
a_x=F_{xf}\cos\delta_f/m+F_{xr}/m
\end{align}

With the assumption of a small tire slip angle, the linear relationship between the lateral tire force and the tire slip angle can be achieved.
\begin{align}
F_{yf}=-C_f\alpha_f, \quad F_{yr}=-C_r\alpha_r
\end{align}
where $C_f$ and $C_r$ are the cornering stiffness of the front and rear tires, respectively. Additionally, the slip angles of the front and rear tires $\alpha_f$ and $\alpha_r$ are given by
\begin{align}
\alpha_f=-\delta_f+(v_y+l_fr)/v_x, \quad \alpha_r=(v_y-l_rr)/v_x
\end{align}
\subsection{Discrete Motion Prediction}
To conduct motion prediction, the single-track vehicle model, Eq. (1) is transformed into a time-varying linear system:
\begin{align}
\dot{x}(t)=A_tx(t)+B_tu(t)
\end{align}
where the time-varying coefficient matrices are derived as
\begin{align}
A_t=\left.\frac{\mathrm{\partial}\Gamma}{\mathrm{\partial}x}\right|_{x_t,u_t},\ B_t=\left.\frac{\mathrm{\partial}\Gamma}{\mathrm{\partial}u}\right|_{x_t,u_t}
\end{align}

And then, Eq. (6) can be discretized as
\begin{align}
\left\{
\begin{array}{lr}
x(k+1)=A_kx(k)+B_ku(k)\\
u(k)=u(k-1)+\Delta{u(k)}\\
\end{array}
\right.
\end{align}
where $x(k)=[v_x(k),v_y(k),r(k),\varphi(k),X(k),Y(k)]^{T}$, $A_k=e^{A_t\Delta{T}}$,\ $B_k=\int_{0}^{\Delta{T}}{e^{A_t\tau}}B_td\tau$, $\Delta{T}$ is the sampling time,  $u(k)=[a_x(k),\delta_f(k)]^{T}$, $\Delta{u(k)}=[\Delta{a_x(k)},\Delta{\delta_f(k)}]^{T}$.

Then, a new state vector is defined to integrate the original state vector and the control input.
\begin{align}
\vartheta(k)=[x(k),u(k-1)]^T
\end{align}

As a result, a new discrete state-space form of Eq. (8) is derived as
\begin{align}
\left\{
\begin{array}{lr}
\vartheta(k+1)=\tilde{A}_k\vartheta(k)+\tilde{B}_k\Delta{u(k)}\\
y(k)=\tilde{C}_k\vartheta(k)\\
\end{array}
\right.
\end{align}
where $\tilde{A}_k=
\left[
\begin{array}{ccc}
A_k & B_k\\
0_{2\times6} & I_2\\
\end{array}
\right]$,
$\tilde{B}_k=
\left[
\begin{array}{ccc}
B_k\\
I_2\\
\end{array}
\right]$, and
$\tilde{C}_k=
\left[
\begin{array}{ccc}
I_6 & 0_{6\times2}\\
\end{array}
\right]$.

Next, the predictive horizon $N_p$  and the control horizon $N_c$ are defined, $N_p>N_c$. At the time step $k$, if the state vector $\vartheta(k)$, the control vector$\Delta{u(k)}$ and coefficient matrices i.e., $\tilde{A}_{p,k}$, $\tilde{B}_{p,k}$ and $\tilde{C}_{p,k}$ are known, the predicted state vectors can be expressed as
\begin{align}
\left\{
\begin{array}{lr}
\vartheta(p+1|k)=\tilde{A}_{p,k}\vartheta(p|k)+\tilde{B}_{p,k}\Delta{u(p|k)}\\
y(p|k)=\tilde{C}_{p,k}\vartheta(p|k)\\
\end{array}
\right.
\end{align}
where $p=k,k+1,\cdot\cdot\cdot,k+N_p-1$.

Assuming that $\tilde{A}_{p,k}=\tilde{A}_k$, $\tilde{B}_{p,k}=\tilde{B}_k$ and $\tilde{C}_{p,k}=\tilde{C}_k$, it yields that
\begin{align}
\begin{array}{lr}
\vartheta(k+1|k)=\tilde{A}_k\vartheta(k|k)+\tilde{B}_k\Delta{u(k|k)}\\
\vartheta(k+2|k)=\tilde{A}_k^2\vartheta(k|k)+\tilde{A}_k\tilde{B}_k\Delta{u(k|k)}+\tilde{B}_k\Delta{u(k+1|k)}\\
\quad \quad\vdots\\
\vartheta(k+N_c|k)=\tilde{A}_k^{N_c}\vartheta(k|k)+\tilde{A}_k^{N_c-1}\tilde{B}_k\Delta{u(k|k)}+\cdot\cdot\cdot\\
\quad \quad\quad\quad\quad\quad\quad+\tilde{B}_k\Delta{u(k+N_c-1|k)}\\
\quad \quad\vdots\\
\vartheta(k+N_p|k)=\tilde{A}_k^{N_p}\vartheta(k|k)+\tilde{A}_k^{N_p-1}\tilde{B}_k\Delta{u(k|k)}+\cdot\cdot\cdot\\
\quad \quad\quad\quad\quad\quad\quad+\tilde{A}_k^{N_p-N_c}\tilde{B}_k\Delta{u(k+N_c-1|k)}\\
\end{array}
\end{align}

Defining the output vector sequence as
\begin{align}
\mathbf{Y}(k)=[y^T(k+1|k),y^T(k+2|k),\cdot\cdot\cdot,y^T(k+N_p|k)]^T
\end{align}

According to Eqs. 12 and 13, the predicted motion output vector sequence $\mathbf{Y}(k)$ is derived as
\begin{align}
\mathbf{Y}(k)=\bar{C}\vartheta(k|k)+\bar{D}\Delta{\mathbf{u}(k)}
\end{align}
where
$\Delta\mathbf{u}(k)=[\Delta{u}^T(k|k),\Delta{u}^T(k+1|k),\cdot\cdot\cdot,\Delta{u}^T(k+N_c-1|k)]^T$,
$\bar{C}=[(\tilde{C}_k\tilde{A}_k)^T,(\tilde{C}_k\tilde{A}_k^2)^T,\cdot\cdot\cdot,(\tilde{C}_k\tilde{A}_k^{N_p})^T]^T$,
$\bar{D}=
\left[
\begin{array}{ccccc}
\tilde{C}_k\tilde{B}_k & 0 & 0 & 0\\
\vdots & \vdots & \vdots & \vdots\\
\tilde{C}_k\tilde{A}_k^{N_c-1}\tilde{B}_k & \cdots & \tilde{C}_k\tilde{A}_k\tilde{B}_k & \tilde{C}_k\tilde{B}_k\\
\vdots & \vdots & \vdots & \vdots\\
\tilde{C}_k\tilde{A}_k^{N_p-1}\tilde{B}_k & \cdots & \tilde{C}_k\tilde{A}_k^{N_p-N_c+1}\tilde{B}_k & \tilde{C}_k\tilde{A}_k^{N_p-N_c}\tilde{B}_k\\
\end{array}
\right]$.
\\

After the above procedures, the motion prediction of CAVs is finished. Based on the motion prediction, the control vector sequence $\Delta\mathbf{u}$ for CAVs can be figured out by solving the cooperative decision-making problem formulated in the following section.

\section{Decision Making using the Coalition Game Approach}
In this section, a coalitional game approach is applied to the decision-making problem formulated for the CAVs at the multi-lane merging zone. Firstly, four typical types of coalition are proposed. Then, the decision-making cost function is defined with consideration of safety, comfort and efficiency. Considering multiple constraints, the predictive decision-making sequence of each coalition is figured out based on MPC.
\subsection{Formulation of the Coalitional Game for CAVs}
Coalitional game is a typical cooperative game, which aims to minimize the coalition's cost via cooperation. The definition of coalitional game is described as follows.

\textbf{Definition 1} [34]: In a coalitional game, the set of all players is denoted by $N=\{1,2,\cdots,n\}$, who seek to form coalitions to reduce costs. Each subset $S$ of $N$ is called a coalition, i.e., $S\in2^N$. If $S$ consists of only one player, it is also regarded as a coalition, i.e., a single player coalition. If $S$ consists of all players, it is called a grand coalition. A coalitional game is defined by a pair $\langle N,U,J\rangle$, where $U$ is a set of decision-making behaviors of players, $J$ is the characteristic function.

\textbf{Remark 1} : In the game theoretic approach, the characteristic function $J$ is usually denoted by a reward function. Each player or coalition aims to maximize the reward value. While, $J$ corresponds to the minimum cost in this paper. The establishment of the cost function is presented in the next subsection.

For any coalition $S$, the corresponding characteristic function is denoted by $J^S(U_S)$, $S\in2^N$. For a single player coalition, the characteristic function is expressed as $J^i(U_i)$, $i\in N$. In the coalitional game, each player can choose to join any coalition according to its own interest. The choice depends on the individual rationality, which is defined as follows.

\textbf{Definition 2} [35]: Based on the principle of fairness, the individual rationality requires that each player in the coalition should obtain a satisfactory cost allocation which is no more than that without joining the coalition, i.e., $Q_i\leq J^i(U_i)$, $\forall i\in N$, where $Q_i$ is the cost allocation of the player $i$, which is allocated by the Shapley method.

According to the Definition 2, the combination and splitting rules of coalitions are summarized. Any collection of disjoint coalitions $S_j$, $S_j\in2^N$, can be combined together to form a single coalition $H$ if and only if
\begin{align}
\begin{array}{lr}
J^H(U_H)\leq\sum\limits_{j=1}^mJ^{S_j}(U_j)\\[1.4ex]
H=S_1\bigcup S_2\cdots\bigcup S_m, \ j=\{1,2,\cdots,m\}\\
\end{array}
\end{align}

Otherwise, the coalition H splits into smaller coalitions.

\textbf{Remark 2} : To define the game players in the merging decision-making problem, only the direct participants and their adjacent passive participants are regarded as players in the game. The lead and following vehicles are not considered. For instance, in Fig. 1, V1 intents to merge into the main road. Hence, V1 is considered as the direct participant in the game. The merging behavior of V1 would directly affect the action of V2. If V2 changes its lane, its lane-change behavior would further affect the behavior of V3, V2 and V3 are seen as the adjacent passive participants, joining in the cooperative merging decision-making game. Therefore, V1, V2 and V3 form the set of game players. The lead and following vehicles of V1, V2 and V3 are considered as the ahead and following game player sets, respectively. The direct interaction and game between the host coalition and its upstream coalition or downstream coalition will be further considered in our future work.

In the multi-lane merging scenario, the coalition of CAVs aims to minimize the cost function of their decision making, which is related to safety, comfort and efficiency performance. The cost function is defined in details in the next subsection. At the multi-lane merging zone, four typical types of coalition are illustrated in Fig. 3. The first type is the single-player coalition, as shown in Fig. 3 (a). Each CAV is unwilling to form a larger coalition and cooperate with others. Hence, the single-player coalition can be seen as a noncooperative game. The second coalition type is a multi-player one, as presented in Fig. 3 (b). In this scenario, V1 and V2 merges into a two-player coalition, and cooperatively address the merging issue. Fig. 3 (c) shows the type of grand coalition, and all vehicles form a large single coalition in the merging scenario. In another words, all the three CAVs participate in the cooperative game. While Fig. 3 (d) presents another coalition form that contains a sub-coalition. With the consideration of the small gap and the same merging demand, V1 and V4 are regarded as one player to join the game and cooperate with others. As a result, V1 and V4 would make a same decision during the game. Generally, the above four types should follow the rules of coalitional game. Therefore, the coalition type may change over time to adapt to different factors and varying merging situations. Although the four types of coalition illustrated in Fig. 3 are not exhaustive, they are the paternal line, which can derive other coalitional types.

\begin{figure}[h]\centering
	\includegraphics[width=8.5cm]{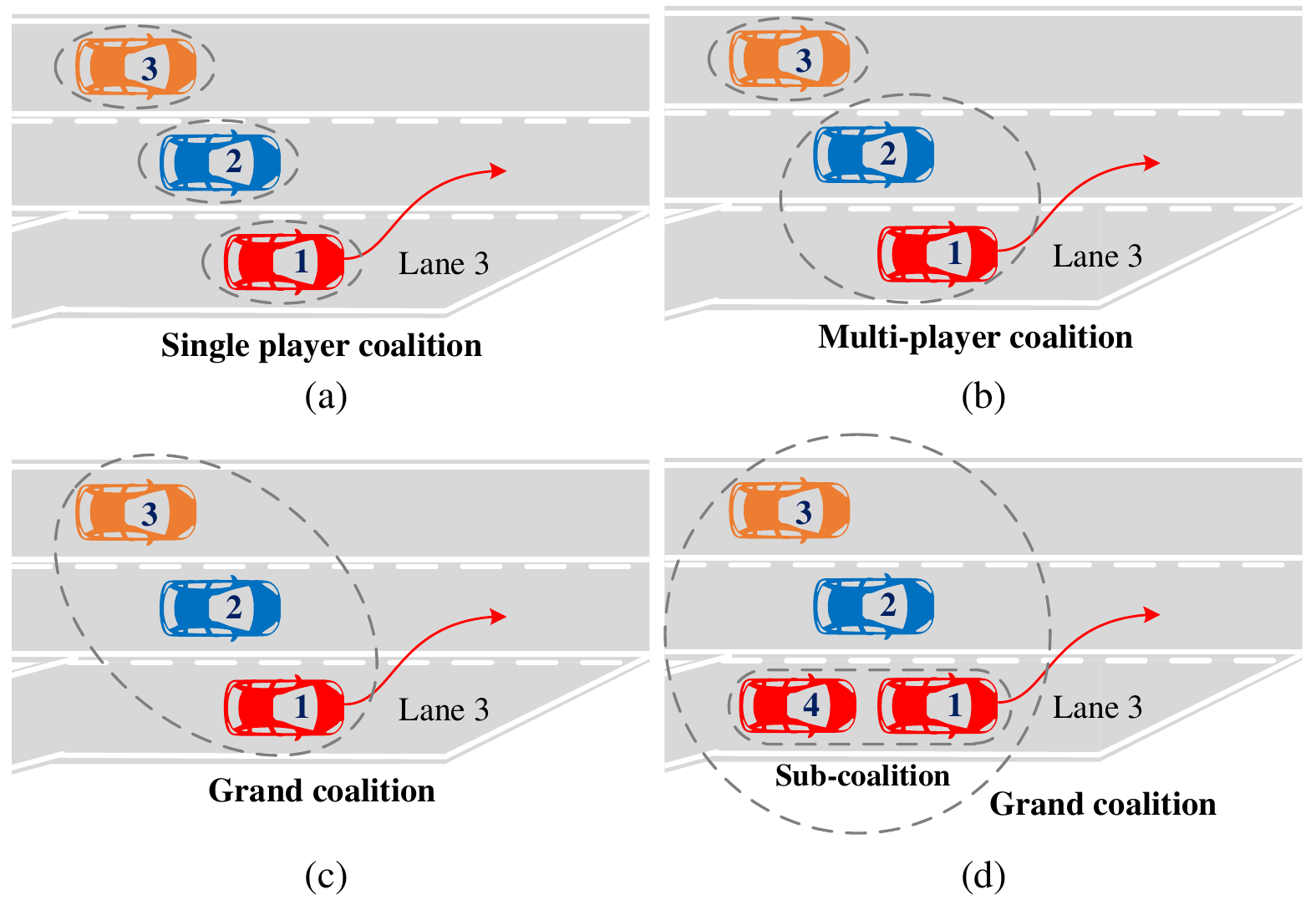}
	\caption{Four types of coalition for CAVs at the multi-lane merging zone: (a) The single player coalition; (b) The multi-player coalition; (c) The grand coalition; (d) The grand coalition with a sub-coalition.}\label{FIG_3}
\end{figure}

\subsection{Cost Function for the Decision Making of an Individual CAV}
In the merging decision-making scenario for CAVs, the final goal of V1 is to merge to the main road, i.e., changing its lane from Lane 3 to Lane 2. V2 is set to be able to accelerate, decelerate or change lane to Lane 1, while V1 can only control its longitudinal motion, i.e., accelerating or decelerating. It can be found that for all the three vehicles at the multi-lane merging zone, their decision making is associated with two driving maneuvers, i.e., acceleration control and lane-change. The acceleration control is related to the driving behavior of the lead vehicle (LV), and the lane-change behavior is affected by the reaction of the neighbor vehicle (NV). Thus, the decision-making process of the host vehicle (HV) Vi is related to the behaviors of LV and NV. Thanks to the V2V technique, the motion state information of each CAV, including acceleration, velocity and position, can be shared among surrounding ones to obtain a better performance of the cooperative decision making.

In the decision-making modeling process of CAVs, three key factors are considered, namely, the safety, comfort and efficiency. Thus, the cost function of the decision consists of three terms. For Vi, the cost function of decision making is expressed as
\begin{align}
J^{Vi}=\omega_s^{Vi}J_s^{Vi}+\omega_c^{Vi}J_c^{Vi}+\omega_e^{Vi}J_e^{Vi}
\end{align}
where $J_s^{Vi}$, $J_c^{Vi}$ and $J_e^{Vi}$ denote the costs of driving safety, ride comfort and travel efficiency, respectively. $\omega_s^{Vi}$, $\omega_c^{Vi}$ and $\omega_e^{Vi}$ are the weighting coefficients, which reflect the driving characteristic of Vi. Referring to [36, 37], the weighting coefficients of the three different driving characteristics are set and listed in Table I.

The cost term on safety  $J_s^{Vi}$ consists of three parts, i.e., the longitudinal safety, lateral safety and lane keeping safety. It can be given by
\begin{align}
\begin{array}{lr} J_s^{Vi}=((\beta^{Vi})^2-1)^2J_{s-log}^{Vi}+(\beta^{Vi})^2J_{s-lat}^{Vi}\\[1.4ex]
\quad \quad \quad  \quad +((\beta^{Vi})^2-1)^2J_{s-lk}^{Vi}+J_{s-lc}^{Vi}
\end{array}
\end{align}
where $J_{s-log}^{Vi}$, $J_{s-lat}^{Vi}$, $J_{s-lk}^{Vi}$ and $J_{s-lc}^{Vi}$ denote the costs of the longitudinal, lateral, lane-keeping and lane-change safety, respectively. $\beta^{Vi}$ is the lane-change behavior of Vi, $\beta^{Vi}\in\{-1,0,1\}:=$ \{\emph{left lane change, lane keeping, right lane change}\}.

\begin{table}[h]
	\renewcommand{\arraystretch}{1.3}
	\caption{Weighting Coefficients of Different Driving Characteristics}
\setlength{\tabcolsep}{6mm}
	\centering
	\label{table_1}
	\resizebox{\columnwidth}{!}{
		\begin{tabular}{l l l l}
			\hline\hline \\[-3mm]
			\multirow{2}{*}{Driving Characteristic} &\multicolumn{3}{c}{Weighting Coefficients} \\
\cline{2-4} & $\omega_s^{Vi}$ & $\omega_c^{Vi}$ & $\omega_e^{Vi}$ \\
\hline
			\multicolumn{1}{c}{Aggressive}  & 0.1 & $ 0.1 $ & 0.8\\
			\multicolumn{1}{c}{Moderate} & 0.5 & 0.3 & 0.2 \\
\multicolumn{1}{c}{Conservative} & 0.7 & 0.2 & 0.1 \\
			\hline\hline
		\end{tabular}
	}
\end{table}

The cost on the longitudinal safety $J_{s-log}^{Vi}$, which is associated with the longitudinal gap and relative velocity with respective to LV, can expressed as
\begin{align}
\begin{array}{lr}
J_{s-log}^{Vi}=\varpi_{v-log}^{Vi}\eta_\sigma^{Vi}(\Delta v_{x,\sigma}^{Vi})^2 +\varpi_{s-log}^{Vi}/[(\Delta s_{\sigma}^{Vi})^2+\varepsilon]\tag{18a}
\end{array}
\end{align}
\begin{align}
\Delta v_{x,\sigma}^{Vi}=v_{x,\sigma}^{LV}-v_{x,\sigma}^{Vi}
\tag{18b}
\end{align}
\begin{align}
\begin{array}{lr}
\Delta s_{\sigma}^{Vi}=[(X_{\sigma}^{LV}-X_{\sigma}^{Vi})^2\\[1.4ex]
\quad \quad \quad  \quad \quad+(Y_{\sigma}^{LV}-Y_{\sigma}^{Vi})^2]^{1/2}-L_{V}
\tag{18c}
\end{array}
\end{align}
\begin{align}
\eta_\sigma^{Vi}=0.5-0.5\mathrm{sgn}(\Delta v_{x,\sigma}^{Vi})
\tag{18d}
\end{align}
where $v_{x,\sigma}^{LV}$ and $v_{x,\sigma}^{Vi}$ denote the longitudinal velocities of LV and Vi, respectively. $(X_{\sigma}^{LV},Y_{\sigma}^{LV})$  and $(X_{\sigma}^{Vi},Y_{\sigma}^{Vi})$ are the positions of LV and Vi, respectively. $\varpi_{v-log}^{Vi}$ and $\varpi_{s-log}^{Vi}$ are the weighting coefficients. $\varepsilon$ is a design parameter of a small value to avoid zero denominator in the calculation. $L_{V}$ is a safety coefficient considering the length of the vehicle. $\sigma$ denotes the lane number labelled from left to right, $\sigma\in\{1,2,3\}$ :=\{\emph{lane 1, lane 2, lane 3}\}. $\eta_\sigma^{Vi}$ is a switch function. If $\Delta v_{x,\sigma}^{Vi}>0$, i.e., $v_{x,\sigma}^{LV}>v_{x,\sigma}^{Vi}$, $\eta_\sigma^{Vi}=0$. As a result, the cost of longitudinal safety is only associated with the relative distance. Otherwise, it is related to both the relative distance and the relative velocity.

The cost on the lateral safety $J_{s-lat}^{Vi}$, which is associated with the relative distance and relative velocity with respective to NV, can be given by
\begin{align}
\begin{array}{lr}
J_{s-lat}^{Vi}=\varpi_{v-lat}^{Vi}\eta_{\sigma+\beta^{Vi}}^{Vi}(\Delta v_{x,\sigma+\beta^{Vi}}^{Vi})^2\\[1.4ex]
\quad \quad \quad  \quad \quad \quad \quad +\varpi_{s-lat}^{Vi}/[(\Delta s_{\sigma+\beta^{Vi}}^{Vi})^2+\varepsilon]
\tag{19a}
\end{array}
\end{align}
\begin{align}
\Delta v_{x,\sigma+\beta^{Vi}}^{Vi}=v_{x,\sigma}^{Vi}-v_{x,\sigma+\beta^{Vi}}^{NV}
\tag{19b}
\end{align}
\begin{align}
\begin{array}{lr}
\Delta s_{\sigma+\beta^{Vi}}^{Vi}=[(X_{\sigma}^{Vi}-X_{\sigma+\beta^{Vi}}^{NV})^2\\[1.4ex]
\quad \quad \quad\quad \quad \quad+(Y_{\sigma}^{Vi}-Y_{\sigma+\beta^{Vi}}^{NV})^2]^{1/2}-L_{V}
\tag{19c}
\end{array}
\end{align}
\begin{align}
\eta_{\sigma+\beta^{Vi}}^{Vi}=0.5-0.5\mathrm{sgn}(\Delta v_{x,\sigma+\beta^{Vi}}^{Vi})
\tag{19d}
\end{align}
where $v_{x,\sigma+\beta^{Vi}}^{NV}$ is the longitudinal velocity of NV. $(X_{\sigma+\beta^{Vi}}^{NV},Y_{\sigma+\beta^{Vi}}^{NV})$ is the position of NV. $\varpi_{v-lat}^{Vi}$ and $\varpi_{s-lat}^{Vi}$  are the weighting coefficients. $\eta_{\sigma+\beta^{Vi}}^{Vi}$ is a switch function. If $\Delta v_{x,\sigma+\beta^{Vi}}^{Vi}>0$, i.e., $v_{x,\sigma}^{Vi}>v_{x,\sigma+\beta^{Vi}}^{NV}$, $\eta_{\sigma+\beta^{Vi}}^{Vi}=0$. As a result, the cost of lateral safety is only correlated to the relative distance. Otherwise, it is related to both the relative distance and the relative velocity.

The cost on the lane keeping safety $J_{s-lk}^{Vi}$ is related to the lateral distance error and the yaw angle error between the predicted position of Vi and the center line of the lane $\sigma$ . It is defined as
\begin{align}
J_{s-lk}^{Vi}=\varpi_{y-lk}^{Vi}(\Delta y_\sigma^{Vi})^2+\varpi_{\varphi-lk}^{Vi}(\Delta \varphi_\sigma^{Vi})^2
\tag{20}
\end{align}
where $\Delta y_\sigma^{Vi}$ and $\Delta \varphi_\sigma^{Vi}$  are the lateral distance error and yaw angle error, $\varpi_{y-lk}^{Vi}$  and $\varpi_{\varphi-lk}^{Vi}$  are the weighting coefficients.

The cost of lane-change safety $J_{s-lc}^{Vi}$ is used to guarantee the driving safety during the lane change process, and it can be calculated based on a potential field model [30].
\begin{align}
J_{s-lc}^{Vi}=\Gamma^{LV}+\Gamma^{NV}
\tag{21a}
\end{align}
\begin{align}
\Gamma^{\jmath}=\hbar^\jmath e^\Psi,\quad (\jmath=LV, NV)
\tag{21b}
\end{align}
\begin{align}
\begin{array}{lr}
\Psi=-\{\frac{\hat{X}^2}{2\sigma_X^2}+\frac{\hat{Y}^2}{2\sigma_Y^2}\}^\varrho+\varsigma v_x^\jmath\Upsilon
\tag{21c}
\end{array}
\end{align}
\begin{align}
\Upsilon=k^\jmath\frac{\hat{X}^2}{2\sigma_X^2}/\sqrt{\frac{\hat{X}^2}{2\sigma_X^2}+\frac{\hat{Y}^2}{2\sigma_Y^2}}
\tag{21d}
\end{align}
\begin{align}
k^\jmath=
&
\left\{
\begin{array}{lr}
-1,\hat{X}<0\\
1,\quad\hat{X}\geq 0\\
\end{array}
\right.
\tag{21e}
\end{align}
\begin{align}
\left[
\begin{array}{ccc}
\hat{X}\\
\hat{Y}\\
\end{array}
\right]=
&
\left[
\begin{array}{ccc}
\cos\varphi^\jmath \quad \sin\varphi^\jmath\\
-\sin\varphi^\jmath \quad \cos\varphi^\jmath\\
\end{array}
\right]
\left[
\begin{array}{ccc}
X-X^\jmath\\
Y-Y^\jmath\\
\end{array}
\right]
\tag{21f}
\end{align}
where $\Gamma^{\jmath}$ denotes the potential field value induced by LV or NV at the position $(X,Y)$. $(X^\jmath,Y^\jmath)$ denotes the position coordinates of LV and NV. $\varphi^\jmath$ and $v_x^\jmath$ denote the yaw angle and velocity of LV and NV. $\sigma_X$ and $\sigma_Y$ are the convergence coefficients. $\hbar$, $\varrho$ and $\varsigma$ are the shape coefficients.

The cost term on ride comfort  $J_{c}^{Vi}$, which is associated with the jerk, is written as
\begin{align}
\begin{array}{lr}
J_{c}^{Vi}=\varpi_{j_x}^{Vi}(j_{x,\sigma}^{Vi})^2+\varpi_{j_y}^{Vi}(j_{y,\sigma}^{Vi})^2
\tag{22}
\end{array}
\end{align}
where $j_{x,\sigma}^{Vi}$ and  $j_{y,\sigma}^{Vi}$ are the longitudinal and lateral jerks of Vi. $\varpi_{j_x}^{Vi}$ and  $\varpi_{j_y}^{Vi}$ are the weighting coefficients.

Additionally, the cost on travel efficiency $J_e^{Vi}$  is designed to be a function of the longitudinal velocity of Vi. It can be expressed by
\begin{align}
J_e^{Vi}=\varpi_e^{Vi}(v_{x,\sigma}^{Vi}-\hat{v}_{x,\sigma}^{Vi})^2
\tag{23a}
\end{align}
\begin{align}
\hat{v}_{x,\sigma}^{Vi}=\min(v_{x,\sigma}^{\mathrm{max}},v_{x,\sigma}^{LV})
\tag{23b}
\end{align}
where $v_{x,\sigma}^{\mathrm{max}}$  is the velocity limit on the lane  $\sigma$, and $\varpi_e^{Vi}$  is the weighting coefficient. In this paper, the min-max normalization is conducted before the set-up of all weighting coefficients.

It should be noted that the name of NV, LV and HV are not fixed. With the change of HV's relative position, the roles of the surrounding vehicles would change accordingly, i.e., the leader-follower topology in the graph would be altered.
\subsection{Constraints of the Decision Making}
In terms of safety, comfort and efficiency, some constraints must be considered in the process of decision making. The safety constraints for Vi are defined as follows.
\begin{align}
|\Delta s_\sigma^{Vi}|\leq\Delta s^{\mathrm{max}},|\Delta y_\sigma^{Vi}|\leq\Delta y^{\mathrm{max}},|\Delta \varphi_\sigma^{Vi}|\leq\Delta \varphi^{\mathrm{max}}
\tag{24}
\end{align}

The constraints for ride comfort are given by
\begin{align}
|j_{x,\sigma}^{Vi}|\leq j_x^{\mathrm{max}},|j_{y,\sigma}^{Vi}|\leq j_y^{\mathrm{max}}
\tag{25}
\end{align}

The constraints for accelerations are defined by
\begin{align}
|a_{x,\sigma}^{Vi}|\leq a_x^{\mathrm{max}},|a_{y,\sigma}^{Vi}|\leq a_y^{\mathrm{max}}
\tag{26}
\end{align}

And the constraint for travel efficiency can be represented as
\begin{align}
|v_{x,\sigma}^{Vi}|\leq v_{x,\sigma}^{\mathrm{max}}
\tag{27}
\end{align}

In addition, the constraint of the curvature trajectory during the lane change process is also considered.
\begin{align}
\frac{|\dot{X}_\sigma^{Vi}\ddot{Y}_\sigma^{Vi}-\ddot{X}_\sigma^{Vi}\dot{Y}_\sigma^{Vi}|}{[(\dot{X}_\sigma^{Vi})^2+(\dot{Y}_\sigma^{Vi})^2]^{3/2}}\leq \frac{1}{R_{\mathrm{min}}}
\tag{28}
\end{align}
\begin{align}
\dot{X}_\sigma^{Vi}=[X_\sigma^{Vi}(k+1)-X_\sigma^{Vi}(k)]/\Delta T
\tag{28a}
\end{align}
\begin{align}
\dot{Y}_\sigma^{Vi}=[Y_\sigma^{Vi}(k+1)-Y_\sigma^{Vi}(k)]/\Delta T
\tag{28b}
\end{align}
\begin{align}
\ddot{X}_\sigma^{Vi}=[X_\sigma^{Vi}(k+2)-2X_\sigma^{Vi}(k+1)+X_\sigma^{Vi}(k)]/\Delta T^2
\tag{28c}
\end{align}
\begin{align}
\ddot{Y}_\sigma^{Vi}=[Y_\sigma^{Vi}(k+2)-2Y_\sigma^{Vi}(k+1)+Y_\sigma^{Vi}(k)]/\Delta T^2
\tag{28d}
\end{align}
where $R_{\mathrm{min}}$ is the minimum turning radius.

Moreover, the constraint of the steering angle of the front wheels is given by
\begin{align}
|\delta_f^{Vi}|\leq\delta_f^{\mathrm{max}},\quad |\Delta\delta_f^{Vi}|\leq\Delta\delta_f^{\mathrm{max}}
\tag{29}
\end{align}

The constraint of $\Delta a_x^{Vi}$  is defined as
\begin{align}
|\Delta a_x^{Vi}|\leq\Delta a_x^{\mathrm{max}}
\tag{30}
\end{align}

In general, the aforementioned constraints for Vi can be expressed in a compact form as
\begin{align}
\begin{array}{lr}
\Phi^{Vi}(\Delta s_\sigma^{Vi},\Delta y_\sigma^{Vi},\Delta \varphi_\sigma^{Vi},a_{x,\sigma}^{Vi},a_{y,\sigma}^{Vi},j_{x,\sigma}^{Vi},\\ [1.4ex]
\quad j_{y,\sigma}^{Vi}, v_{x,\sigma}^{Vi},X_\sigma^{Vi},Y_\sigma^{Vi},\delta_f^{Vi},\Delta\delta_f^{Vi},\Delta a_x^{Vi})
\tag{31}
\end{array}
\end{align}

\subsection{Decision Making with the Coalitional Game Approach}
Benefiting from the prediction information, the decision-making algorithm based on the motion prediction of CAVs is able to provide more accurate and reliable decisions for the low-level motion planning and control modules. In this section, the MPC method is adopted and applied in the optimization problem formulated for decision-making.

Based on the motion prediction algorithm described in Section III, the cost function sequence of Vi at the time step k can be derived as
\begin{align}
J^{Vi}(k+1|k),J^{Vi}(k+2|k),\cdots,J^{Vi}(k+N_p|k)
\tag{32}
\end{align}

Then the decision-making sequence of Vi can be given by
\begin{align}
\hat{u}^{Vi}(k|k),\hat{u}^{Vi}(k+1|k),\cdots,\hat{u}^{Vi}(k+N_c-1|k)
\tag{33}
\end{align}
where $\hat{u}^{Vi}(q|k)=[\Delta a_x^{Vi}(q|k),\Delta \delta_f^{Vi}(q|k),\beta ^{Vi}(q|k)]^T$, $q=k,k+1,\cdots,k+N_c-1$.

Besides, the characteristic function of Vi for decision making is defined as
\begin{align}
\Lambda^{Vi}=\sum\limits_{p=k+1}^{k+N_p}||J^{Vi}(p|k)||_Q^2+\sum\limits_{q=k}^{k+N_c-1}||\hat{u}^{Vi}(q|k)||_R^2
\tag{34}
\end{align}
where $Q$ and $R$ are the weighting matrices.

According to the aforementioned four coalition types of the CAVs at the multi-lane merging zone, four different decision-making strategies are designed as follows.

(1) The single player coalition:

This coalitional game consists of three single player coalitions, i.e.,  $S_1=\{V1\}$, $S_2=\{V2\}$, $S_3=\{V3\}$, as shown in Fig. 3 (a). The decision-making sequences of the three coalitions are derived as
\begin{align}
(\Delta a_x^{V1*},\Delta \delta_f^{V1*},\beta^{V1*})=\arg\min\Lambda^{V1}
\tag{35a}
\end{align}
\begin{align}
(\Delta a_x^{V2*},\Delta \delta_f^{V2*},\beta^{V2*})=\arg\min\Lambda^{V2}
\tag{35b}
\end{align}
\begin{align}
\Delta a_x^{V3*}=\arg\min\Lambda^{V3}
\tag{35c}
\end{align}
s.t. $\Phi^{V1}\leq0$, $\Phi^{V2}\leq0$, $\Phi^{V3}\leq0$, $\beta^{V1}(\beta^{V1}+1)=0$, $\beta^{V2}(\beta^{V2}+1)=0$. \\
where $\Delta a_x^{Vi*},\Delta \delta_f^{Vi*}$ and $\beta^{Vi*}$  denote the optimal decision-making sequence of Vi.

(2) The multi-player coalition:

This type of the coalitional game consists of two coalitions, i.e., $S_1=\{V1, V2\}$, $S_2=\{ V3\}$, as shown in Fig. 3 (b). The decision-making sequences of the two coalitions are expressed as
\begin{align}
\begin{array}{lr}
(\Delta a_x^{V1*},\Delta \delta_f^{V1*},\beta^{V1*},\Delta a_x^{V2*},\Delta \delta_f^{V2*},\beta^{V2*})\\[1.4ex]
\quad =\arg\min[\Lambda^{V1}+\Lambda^{V2}]
\tag{36a}
\end{array}
\end{align}
\begin{align}
\Delta a_x^{V3*}=\arg\min\Lambda^{V3}
\tag{36b}
\end{align}

s.t. $\Phi^{V1}\leq0$, $\Phi^{V2}\leq0$, $\Phi^{V3}\leq0$, $\beta^{V1}(\beta^{V1}+1)=0$, $\beta^{V2}(\beta^{V2}+1)=0$.

(3) The grand coalition:

This coalitional game concludes all three target CAVs, which forms a grand single coalition, i.e., $S_1=\{V1, V2, V3\}$, as shown in Fig. 3 (c). The decision-making sequence can be described as
\begin{align}
\begin{array}{lr}
(\Delta a_x^{V1*},\Delta \delta_f^{V1*},\beta^{V1*},\Delta a_x^{V2*},\Delta \delta_f^{V2*},\beta^{V2*},\\[1.4ex]
\quad \Delta a_x^{V3*})=\arg\min[\Lambda^{V1}+\Lambda^{V2}+\Lambda^{V3}]
\tag{37}
\end{array}
\end{align}
s.t. $\Phi^{V1}\leq0$, $\Phi^{V2}\leq0$, $\Phi^{V3}\leq0$, $\beta^{V1}(\beta^{V1}+1)=0$, $\beta^{V2}(\beta^{V2}+1)=0$.

(4) The grand coalition including a sub-coalition:

This coalitional game includes all four target four CAVs, i.e., $S_1=\{\{V1, V4\}, V2, V3\}$, as shown in Fig. 3 (d). While, to simplify the decision-making process, V1 and V4 are regarded as one sub-coalition in $S_1$. In this coalitional game, the decision-making behaviors of V1 and V4 are the same, but there might be a time delay existing between the two vehicles. The decision-making sequence of this coalition can be derived as
\begin{align}
\begin{array}{lr}
(\Delta a_x^{V1*},\Delta \delta_f^{V1*},\beta^{V1*},\Delta a_x^{V2*},\Delta \delta_f^{V2*},\beta^{V2*},\\[1.4ex]
\quad \Delta a_x^{V3*})=\arg\min[\Lambda^{V1}+\Lambda^{V2}+\Lambda^{V3}]
\tag{38a}
\end{array}
\end{align}
\begin{align}
\begin{array}{lr}
\Delta a_x^{V4*}(k)=\Delta a_x^{V1*}(k-\tau)\\[1.4ex] \Delta \delta_f^{V4*}(k)=\Delta \delta_f^{V1*}(k-\tau)\\[1.4ex]
\beta^{V4*}(k)=\beta^{V1*}(k-\tau)
\tag{38b}
\end{array}
\end{align}
s.t. $\Phi^{V1}\leq0$, $\Phi^{V2}\leq0$, $\Phi^{V3}\leq0$, $\beta^{V1}(\beta^{V1}+1)=0$, $\beta^{V2}(\beta^{V2}+1)=0$.\\
where $\tau$  denotes the decision-making delay of V4 with respect to V1. It is related to the velocity of V4 and the gap between V1 and V4. $\tau=\Delta s_\sigma^{V4}/v_{x,\sigma}^{V4}/\Delta T$.

\begin{algorithm}[h]
\caption{Decision-making algorithm for $Vi$ at the multi-lane merging zone.}
\begin{algorithmic}[1]
\STATE Input the motion information of $Vi$ (on the Lane 3);
\STATE Input the motion information of $NVs$ and $LV$ for $Vi$. $Vj$ and $Vk$ are the $NVs$ for $Vi$ on the Lane 2 and Lane 1;
\FOR{$i=1:n$}

\STATE Sub-coalition formation with Algorithm 2;
\STATE Coalition $S=\{Si, Sj, Sk\}$;
\FOR{$\xi=1:length(S)$}
\IF {$\ Q_{S(\xi)}>J^{S(\xi)}(U_{S(\xi)})$}
\STATE $S(\xi)$ breaks away from $S$;
\ELSE
\STATE $S(\xi)$ remains in $S$;
\ENDIF
\ENDFOR

\IF {only one sub-coalition, $S(\xi)$, breaks away}
\IF {$S(\xi)$=$Si$}
\STATE $S$ splits into \{$Si$\}, \{$Sj$, $Sk$\};
\ELSE
\IF {$S(\xi)$=$Sj$}
\STATE $S$ splits into \{$Sj$\}, \{$Si$, $Sk$\};
\ELSE
\STATE $S$ splits into \{$Sk$\}, \{$Si$, $Sj$\};
\ENDIF
\ENDIF
\ELSE
\IF {Two or more sub-coalitions break away}
\STATE $S$ splits into \{$Si$\}, \{$Sj$\}, \{$Sk$\};
\ELSE
\STATE No splitting, $S=\{Si, Sj, Sk\}$;
\ENDIF
\ENDIF
\STATE Optimization with Eq.35-Eq.38;
\STATE Output the decision-making results.
\ENDFOR
\end{algorithmic}
\end{algorithm}

\begin{algorithm}[t]
\caption{Algorithm for sub-coalition formation.}
\begin{algorithmic}[1]
\STATE Coalition $Si=\{Vi\}$;
\FOR{$\zeta=i+1:n$}
\IF {$\Delta s^{V\zeta}<\Delta s_0$ and $\omega^{V\zeta}=\omega^{V\zeta-1}$}
\STATE $Si\leftarrow add \quad V\zeta$
\ELSE
\STATE Break;
\ENDIF
\ENDFOR

\STATE Coalition $Sj=\{Vj\}$;
\FOR{$\eta=j+1:p$}
\IF {$\Delta s^{V\eta}<\Delta s_0$ and $\omega^{V\eta}=\omega^{V\eta-1}$}
\STATE $Sj\leftarrow add \quad V\eta$
\ELSE
\STATE Break;
\ENDIF
\ENDFOR

\STATE Coalition $Sk=\{Vk\}$;
\FOR{$\gamma=k+1:q$}
\IF {$\Delta s^{V\gamma}<\Delta s_0$ and $\omega^{V\gamma}=\omega^{V\gamma-1}$}
\STATE $Sk\leftarrow add \quad V\gamma$
\ELSE
\STATE Break;
\ENDIF
\ENDFOR
\STATE Output $Si$, $Sj$, $Sk$.
\end{algorithmic}
\end{algorithm}

The above two algorithms are proposed to describe the formation principle of the coalitions for the merging decision making. According to Algorithm 1, firstly, once the vehicle moves into the merging zone, it will be assigned with a number. $Vi$ in Lane 3 wants to merge into the Lane 2, so the NVs of $Vi$ in Lane 1 and Lane 2 are numbered by $Vk$ and $Vj$, respectively. Then, the sub-coalition formation is done based on Algorithm 2. As a result, three sub-coalitions are created in Lane 1, Lane 2 and Lane3, respectively, denoted by $Si$, $Sj$, $Sk$. To find the optimal coalition formation, the grand coalition $S$ is created at first, i.e., $S=\{ Si, Sj, Sk \}$. Then, the sub-coalitions choose to either break away from $S$ or remain in $S$ according to the principle in Definition 2. As a result, the grand coalition $S$ will split into different coalition types, which are covered by the proposed four types of coalitions. Finally, the decision-making results can be generated according to the optimization problem, which is formulated by Eqs. 35 to 38.

It should be noted that the proposed decision-making algorithm is not limited to the scenario with a maximum number of four vehicles, and it can be expanded to scenarios with more vehicles. According to Algorithm 2, three or more vehicles may construct a sub-coalition on the main road or in the on-ramp lane. Then, the optimal coalition formation can be derived according to Algorithm 1. Finally, the decision-making sequence of each coalition can be figured out accordingly.

In this paper, the game-based decision making issue is finally transformed into a closed-loop iterative optimization process with multi-constraints [37, 38], which is solved with the efficient evolutionary algorithm based on convex optimization theory [39]. The Nash equilibrium solution may not be unique all the time, but its existence can be guaranteed.

\section{Testing, Validation and Discussion}
In this section, two testing cases are designed and tested to verify the feasibility and effectiveness of the proposed cooperative decision-making algorithms. All the driving scenarios are established and implemented on the MATLAB/Simulink platform.
Table II shows the parameter settings of the decision-making algorithm in the driving scenarios [40,41].
\begin{table}[!h]
	\renewcommand{\arraystretch}{1.3}
	\caption{Parameters of the Decision-making Algorithm}
\setlength{\tabcolsep}{6mm}
	\centering
	\label{table_2}
	\resizebox{\columnwidth}{!}{
		\begin{tabular}{c c | c c}
			\hline\hline \\[-4mm]
            Parameter & Value & Parameter & Value \\
\hline
			$\Delta s^{\mathrm{max}}$/ $\mathrm{(m)}$  & 0.8 & $j_y^{\mathrm{max}} $/ $\mathrm{(m/s^3)}$ & 2\\
           $\Delta y^{\mathrm{max}}$/ $\mathrm{(m)}$  & 0.2 & $v_{x,\sigma}^{\mathrm{max}}$/ $\mathrm{(m/s)}$ & 30\\
           $\Delta \varphi^{\mathrm{max}}$/ $\mathrm{(deg)}$  & 2 & $R_\mathrm{min} $/ $\mathrm{(m)}$ & 8\\
           $a_x^{\mathrm{max}}$/ $\mathrm{(m/s^2)}$ & 4 & $\delta_f^{\mathrm{max}} $/ $\mathrm{(deg)}$ & 30\\
           $a_y^{\mathrm{max}}$/ $\mathrm{(m/s^2)}$& 4 & $\Delta\delta_f^{\mathrm{max}} $/ $\mathrm{(deg)}$ & 0.3\\
           $j_x^{\mathrm{max}} $/ $\mathrm{(m/s^3)}$ & 2 & $\Delta a_x^{\mathrm{max}} $/ $\mathrm{(m/s^2)}$ & 0.1\\
           $N_p $ & 5 & $N_c$ & 2\\
			\hline\hline
		\end{tabular}
	}
\end{table}

\subsection{Case Study 1}
The scenario of case study 1 is constructed with three CAVs. Fig. 4 shows the conditions of a single player coalition and a grand coalition, which are the two extreme coalitional forms. Specifically, the single player coalition corresponds to a noncooperative game. While, the grand coalition results in a cooperative game including all three players. In this case, the comparative study of the single player coalition and the grand coalition is conducted. Additionally, the effects of different driving characteristics on the coalition formation are analyzed. The initial position coordinates of V1, V2, V3, and the LVs of V1, V2 and V3 are set as (12, -4), (10, 0), (8, 4), (62, -4), (70, 0) and (68, 4), respectively. In addition, the initial longitudinal velocities of V1, V2, V3, and the LVs of V1, V2 and V3 are set to be 18 m/s, 19 m/s, 20 m/s, 26 m/s, 26 m/s, 26 m/s, respectively.

\begin{figure}[!t]\centering
	\includegraphics[width=8.5cm]{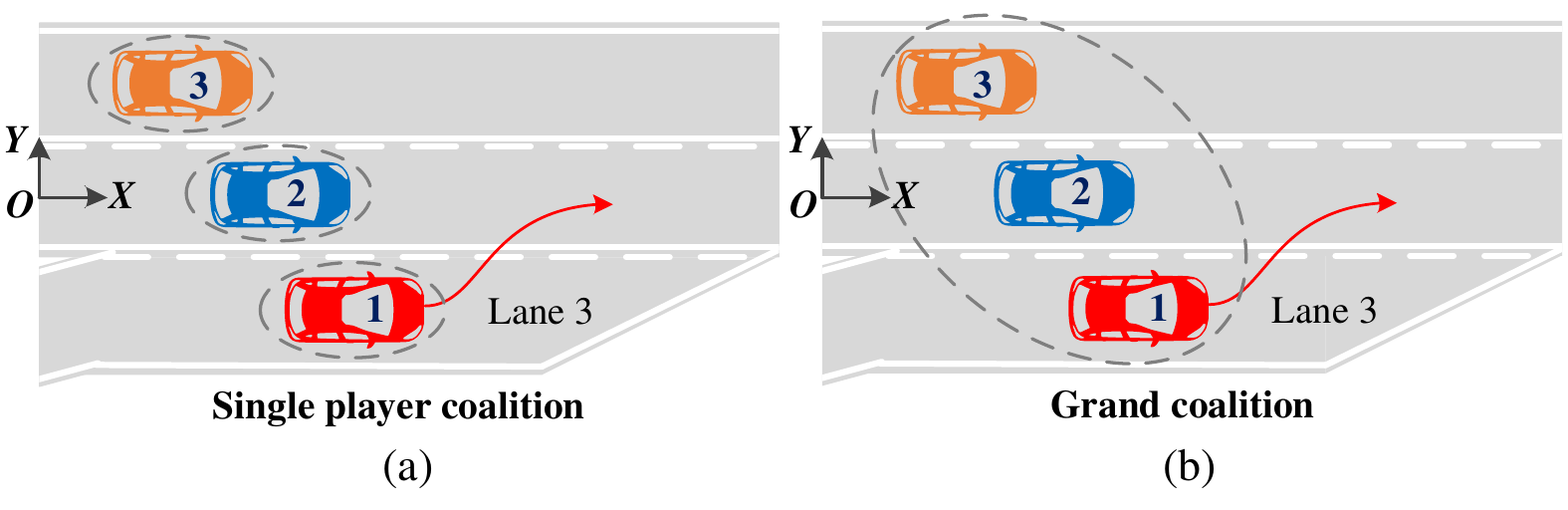}
	\caption{Comparison of the two coalitional types considering different driving characteristics of CAVs: (a) The single player coalition; (b) The grand coalition.}\label{FIG_4}
\end{figure}

Considering different driving characteristics of the CAVs, two scenarios are studied in this case. In Scenario A, the driving characteristics of all the three CAVs are moderate. In Scenario B, the driving characteristics of V2 and V3 are moderate, and V1 is set to be an aggressive driving mode.

The decision-making results of these two scenarios are illustrated in Fig. 5. It can be found that the merging decision of the single player coalition (Coalition 1) is made earlier than that of the grand coalition (Coalition 2). For the single player coalition, each CAV aims to minimize their own cost. While, the grand coalition pursues the minimization of the entire coalitional cost, i.e., the sum cost of all CAVs, rather than the individual one. As a result, the different decision-making results are generated. The detailed testing results are listed in Table III. In Scenario A, the sum cost value of the grand coalition is smaller than that of the single player coalition. Moreover, the cost value of each CAV in the grand coalition is smaller than that in the single player coalition, showing the superiority of the grand coalition. However, in Scenario B, the cost value of V1 in the grand coalition is larger than that in the single player coalition. The only difference between Scenarios A and B is the driving characteristic settings of V1. In Scenario B, the driving characteristic of V1 is aggressive, which is different from other CAVs. Specifically, the weighting coefficients of the cost function for V1 are set to be far different from others.

The testing results of the longitudinal paths and velocities are depicted in Figs. 6 and 7, respectively. As reflected in Fig. 7 (b), the velocity of V1 in the grand coalition is smaller than that in the single player coalition, indicating that the cost on the travel efficiency of V1 increases in the grand coalition. Due to the different driving characteristic of V1, it leads to the increasing costs for individuals in the grand coalition. In other words, the grand coalition is not the optimal choice for V1 in Scenario B. If V1 chooses to join the grand coalition, it would need to bear an additional on its own. However, this is against the Definition 2 of the coalitional rules. Thus, as the grand coalition is not the optimal choice in Scenario B for V1, it would not join the grand coalition.

Additionally, Fig. 8 shows the computational time of the proposed algorithm in Case 1. The mean value of each step is about 0.06s. For real-time experiments in the future, the computational efficiency can be further improved with efficient solver and better computing platforms.

\begin{figure}[!t]\centering
	\includegraphics[width=8.5cm]{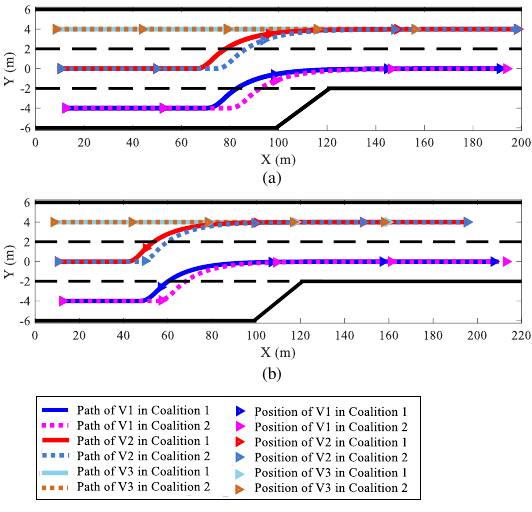}
	\caption{Decision-making results of the two scenarios considering different driving characteristics of CAVs in Case 1: (a) Scenario A; (b) Scenario B.}\label{FIG_5}
\end{figure}

\begin{figure}[!t]\centering
	\includegraphics[width=8.5cm]{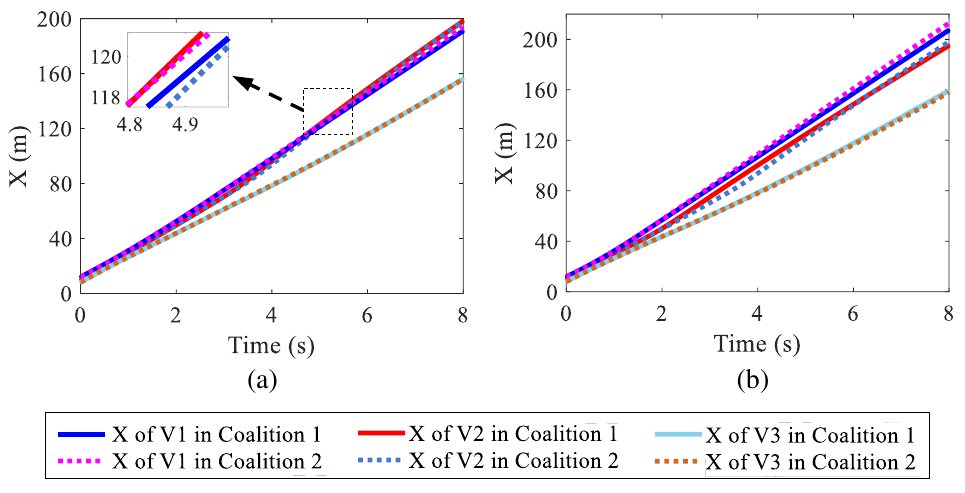}
	\caption{Longitudinal paths of CAVs in the two scenarios considering different driving characteristics in Case 1: (a) Scenario A; (b) Scenario B.}\label{FIG_6}
\end{figure}

\begin{table}[!t]
	\renewcommand{\arraystretch}{1.3}
	\caption{Cost Values of CAVs in Case 1 with Different Scenario Settings}
\setlength{\tabcolsep}{3mm}
	\centering
	\label{table_3}
	\resizebox{\columnwidth}{!}{
		\begin{tabular}{c c c c c}
			\hline\hline \\[-3mm]
			\multirow{2}{*}{Cost Values (RMS)} & \multicolumn{2}{c}{Scenario A} & \multicolumn{2}{c}{Scenario B}\\
\cline{2-5} & \makecell [c] {Single player \\coalition} & \makecell [c] {Grand \\ coalition} & \makecell [c] {Single player \\coalition} & \makecell [c] {Grand \\ coalition}\\
\hline
			\multicolumn{1}{c}{V1} & 62598 & 62196 & 71527 & 72903 \\
			\multicolumn{1}{c}{V2} & 55560 & 55086 & 68933 & 67071  \\
            \multicolumn{1}{c}{V3} & 48521 &  47983  & 65914 & 64297 \\
            \multicolumn{1}{c}{Sum} & 166679 & 165265 & 206374 & 204271 \\
			\hline\hline
		\end{tabular}
	}
\end{table}

\begin{figure}[!h]\centering
	\includegraphics[width=8.5cm]{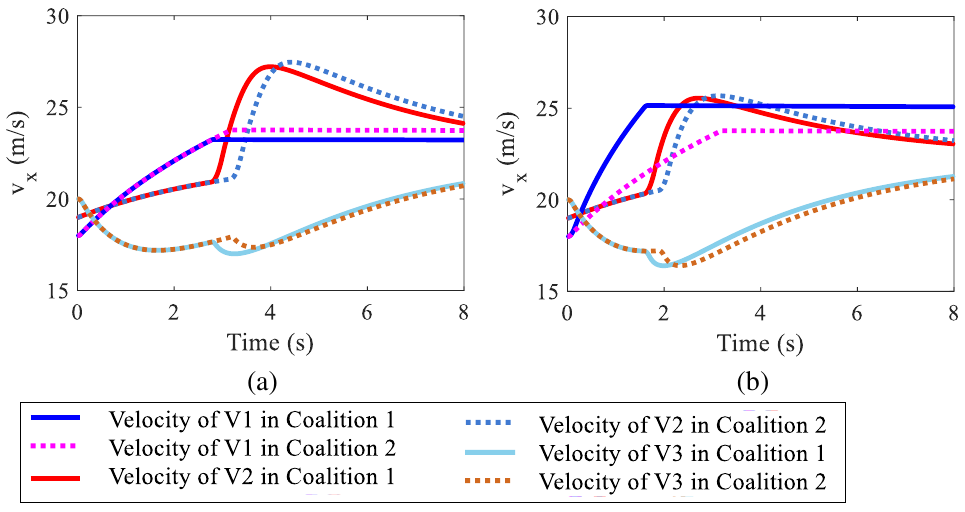}
	\caption{Velocities of CAVs in the two scenarios considering different driving characteristics in Case 1: (a) Scenario A; (b) Scenario B.}\label{FIG_7}
\end{figure}

\begin{figure}[!h]\centering
	\includegraphics[width=8.5cm]{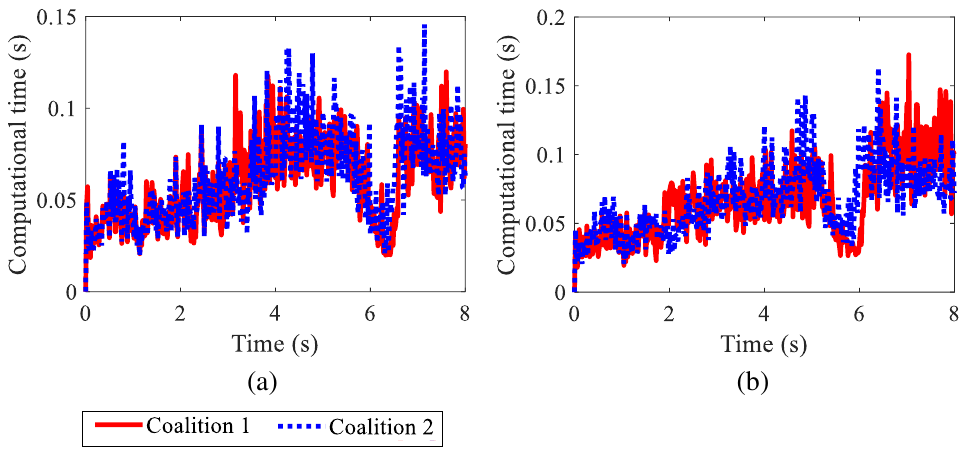}
	\caption{Computational time in Case 1: (a) Scenario A; (b) Scenario B.}\label{FIG_8}
\end{figure}

According to the testing results and analysis of Case 1, some conclusions are drawn. First, the grand coalition can decrease the sum cost of the whole group. Thus, it is beneficial to the traffic system, from the high-level traffic management perspective. When all CAVs are with a same driving characteristic, the grand coalition, which is able to decrease both the individual cost and the overall cost, can be seen as a global optimal solution. However, if the driving characteristics are different among CAVs, this grand coalition may not be optimal for individuals any more, as the optimal coalitional type should follow the previously mentioned Definition 2. This is further investigated and studied in the following Case 2.

\subsection{Case Study 2}
This case mainly studies the effects of driving characteristics on the coalition formation. As Fig. 9 shows, five CAVs are considered here. Considering different driving characteristics, three scenarios are defined in this case. In Scenario A, the driving modes of all four CAVs are set to be moderate. In Scenario B, V1, V2 and V4 are moderate, while V3 and V5 are defined as aggressive ones. In Scenario C, V2 is conservative, while V1, V3, V4 and V5 are moderate driving.

\begin{figure}[t]\centering
	\includegraphics[width=4.5cm]{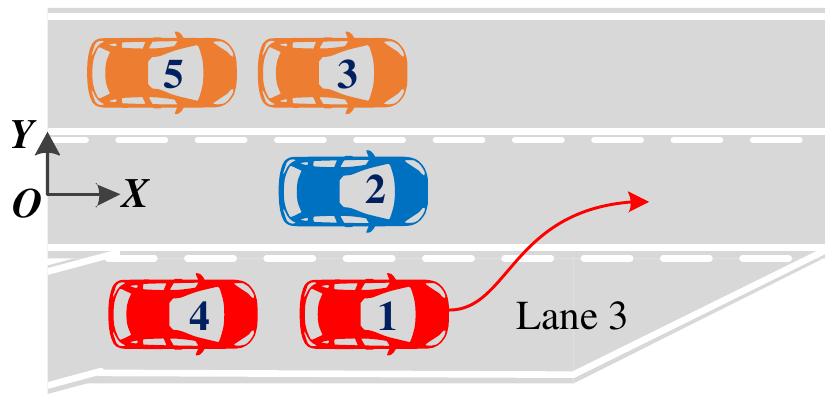}
	\caption{The decision making of four CAVs at the multi-lane merging zone.}\label{FIG_9}
\end{figure}

The initial position coordinates of V1, V2, V3, V4, V5 and the LVs of V1, V2 and V3 are set as (18, -4), (10, 0), (8, 4), (12, -4), (2, 4), (62, -4), (70, 0) and (68, 4), respectively. The initial longitudinal velocities of V1, V2, V3, V4, V5 and the LVs of V1, V2 and V3 are set as 20 m/s, 22 m/s, 16 m/s, 20 m/s, 16 m/s, 28 m/s, 28 m/s, 22 m/s, separately. The testing results are displayed in Figs. 10-14.

\begin{figure}[!t]\centering
	\includegraphics[width=8.5cm]{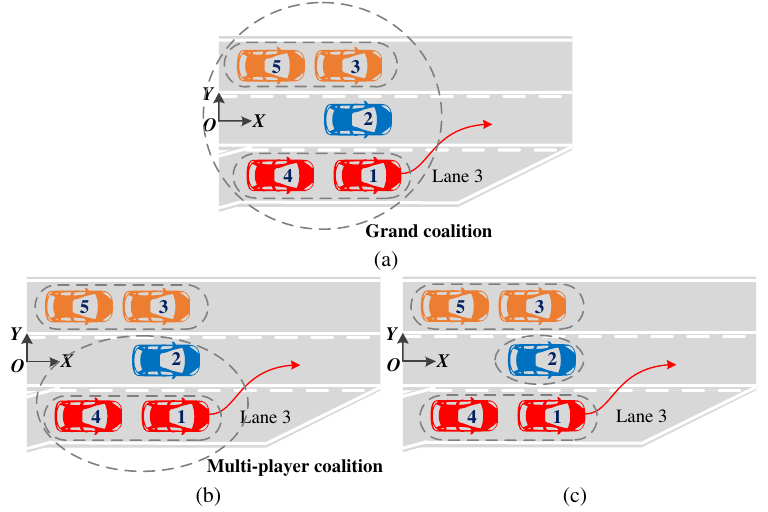}
	\caption{The coalition formation results of CAVs in Case 2: (a) Scenario A; (b) Scenario B; (c) Scenario C.}\label{FIG_10}
\end{figure}

\begin{figure}[!t]\centering
	\includegraphics[width=8.5cm]{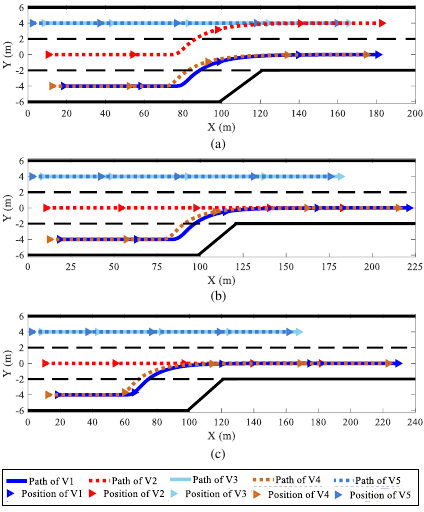}
	\caption{Decision-making results of the three scenarios in Case 2: (a) Scenario A; (b) Scenario B; (c) Scenario C.}\label{FIG_11}
\end{figure}

In Scenario A, all CAVs join into a grand coalition, i.e., $S_1=\{V1, V2, V3, V4, V5\}$, in which V1 and V4 form a sub-coalition, and V3 and V5 form another sub-coalition as well. $S_1\Rightarrow\{\{V1,V4\}, V2, \{V3,V5\}\}$. Since the initial velocity of V2 is larger than that of V1, thus during the cooperative decision-making process, on one hand V2 slows down to create a safer distance for V1 and V4 to cut in, and on the other hand, V2 changes its lane to provide a larger merging space. In Scenario B, since V3 and V5 are both with an aggressive characteristic, their sub-coalition breaks away from the grand coalition. As a result, V1, V2 and V4 form another multi-player coalition, i.e., $S_1=\{\{V1, V4\}, V2\}$ , $S_2=\{V3, V5\}$. Due to the defined aggressive driving characteristics, V3 does not give way to V2. As a result, V2 can only slow down to accommodate the merging behaviors of V1 and V4. Meanwhile, to ensure a larger safe distance, V1 and V4 have to increase their velocities. In addition, Fig. 14 shows the computational time of the proposed algorithm in Case 2. The mean value of each step is about 0.07s, indicating its good potential in real time applications.

\begin{figure}[!t]\centering
	\includegraphics[width=8.5cm]{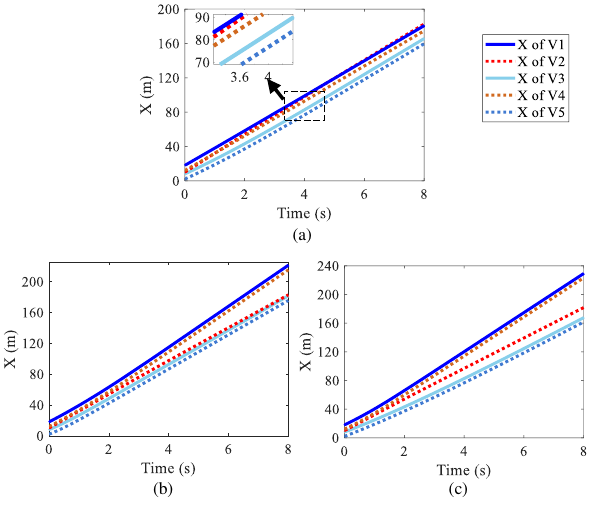}
	\caption{Longitudinal paths of CAVs under the three scenarios in Case 2: (a) Scenario A; (b) Scenario B; (c) Scenario C.}\label{FIG_12}
\end{figure}

\begin{figure}[!t]\centering
	\includegraphics[width=8.5cm]{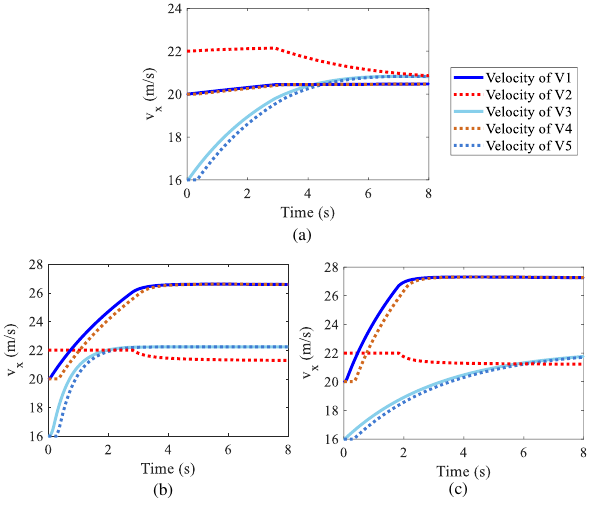}
	\caption{Velocities of CAVs under the three scenarios in Case 2: (a) Scenario A; (b) Scenario B; (c) Scenario C.}\label{FIG_13}
\end{figure}

\begin{figure}[!h]\centering
	\includegraphics[width=4.5cm]{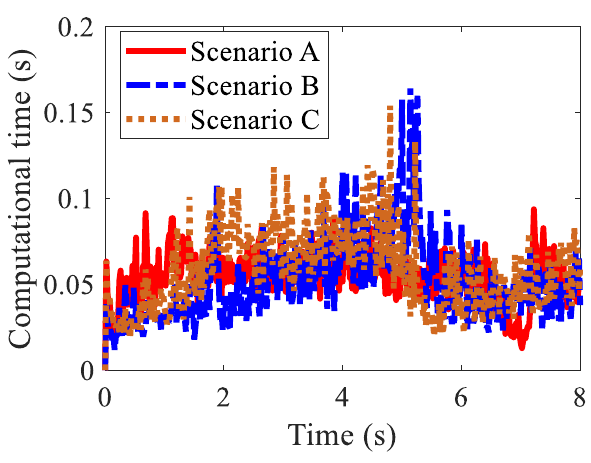}
	\caption{Computational time in Case 2.}\label{FIG_14}
\end{figure}

In Scenario C, V1 and V4 combine as a multi-player coalition, and V2 forms a single player coalitions, while V3 and V5 combine as a multi-player coalition, i.e., $S_1=\{V1, V4\}$, $S_2=\{V2\}$, $S_3=\{ V3, V5\}$ . Although the driving characteristic of V3 is moderate, V2 can only decrease its speed in advance to assist the merging of V1 and V4, rather than changing its lane. This is due to the characteristics of the conservative driving behaviours.

In general, as reflected by the above testing results, different driving characteristics would affect the coalition formations, which is consistent with the coalition formation rule of Definition 2. This is the consideration towards human-like and personalized automated driving. From the system's perspective, the grand coalition is indeed an optimal solution without considering the individual differences. Additionally, if the adjacent CAVs in the same lane have the same driving characteristic and objectives, e.g., merging, they can form a sub-coalition to simplify the game issue.

Additionally, we can find that the proposed decision-making framework is different from either the single vehicle decision making or the centralized control approach. For single vehicle decision making, only the benefits of individuals are considered, while the holistic performance of the traffic system is neglected. For the centralized control approach, it mainly cares about the overall safety and capacity of the entire transportation system, while the personalized demands of individual vehicles are usually neglected. In this work, with the coalitional game approach, the overall performance of the transportation system as well as the different demands of individual vehicles can be coordinated simultaneously through the cooperative decision making.

\section{Conclusion}
This paper presents a cooperative decision-making approach to deal with the multi-lane merging problem for CAVs. To further advance the decision-making algorithm, the motion prediction module is designed based on the vehicle dynamics model. In the decision-making algorithm, the cost function and constraints are defined with consideration of different driving characteristics of CAVs, which are associated with the performances of safety, comfort and efficiency. Furthermore, four typical coalition types are proposed to address the cooperative decision-making issue for CAVs. Under different scenarios and driving characteristics of CAVs, different coalitions are expected to be formed. To evaluate the proposed cooperative decision-making approach, two testing cases are designed and investigated considering different driving characteristics of CAVs. Based on the testing results, the developed coalitional game approach is able to make feasible and reasonable decisions for CAVs at the multi-lane merging zone, and the resultant different coalition formations are adaptive to various driving scenarios. The testing results reflect that the proposed approach is capable to ensure the safety of the traffic system at the complex merging zone, and simultaneously meet personalized driving demands of individuals within the area, demonstrating its feasibility and effectiveness.

Our future work will focus on the decision-making issues with more complex traffic scenarios that includes the human-driven vehicles and CAVs.

\end{document}